# First human trials of a dry electrophysiology sensor using a carbon nanotube array interface

G. Ruffini[1], S. Dunne[1], L. Fuentemilla[2], C. Grau[2], E. Farrés[1], J. Marco-Pallarés[1,2], P. C. P. Watts[3], S. R. P. Silva[3]

(1) Starlab, Ed. de l'Observatori Fabra, C. de l'Observatori s/n, 08035 Barcelona, Spain
(2) Neurodynamics Laboratory, Psychiatry and Clinical Psychobiology Department, University of Barcelona, Spain
(3) Nanoelectronics Centre, Advanced Technology Institute, University of Surrey, Guilford, UK

**Fatigue, sleepiness and disturbed sleep are increasingly important factors affecting health and safety in modern society and there is considerable interest in developing technologies for routine ambulatory monitoring of related physiological indicators [1,2]. Electrophysiology, the measurement of electrical activity of biological origin, is a key technique for the measurement of physiological parameters in several applications, but it has been traditionally difficult to develop sensors for measurements outside the laboratory or clinic with the required quality and robustness [3,4]. This is in part due to the fact that electrodes used for high quality low amplitude measurements (such as EEG) require skin preparation and the use of electrolytic gel, resulting in longer preparation times (up to several minutes per electrode) and long stabilization times (diffusion of the electrolytic gel into the skin). In this paper we report the results from the first series of human trials with a new electrophysiology sensor using multiwalled carbon nanotube arrays (MWCNTs) [5] whose design goal is penetration of the outer layer of the skin and improved electrical contact [6]. These arrays were grown on highly doped silicon substrates using plasma enhanced chemical vapor deposition of acetylene over an iron catalyst and mounted on a back end providing amplification. The trials, which have included traditional protocols for the analysis of the electrical activity of the brain—spontaneous EEG and Event Related Potentials (ERPs)—indicate performance**

**on a par with state of the art research-oriented wet electrodes. No side effects have been observed six months after the tests, and the subject did not report any pain or unusual sensations upon application of the electrode.**

The aim of the tested design was to eliminate skin preparation and gel application requirements in order to reduce noise while improving wearability [6]. The key aspect of this prototype is the electrode-skin interface, which is provided by a large number of MWCNTs forming a brush-like structure — see Figures 1 and 2. Due to the excellent mechanical (high modulus of elasticity, tensile strength [7]) and electrical properties [8] of MWCNTs, this nano-structure design should provide a robust and stable low noise electrical interface of low impedance while barely penetrating the outer skin layer—the *Stratum Corneum*—resulting in a comfortable and pain free interface. In addition, shallow penetration, together with the small diameter of CNTs, should result in a lower infection risk as compared to more invasive approaches using microtechnology.

The mechanical interaction of MWCNTs and skin is poorly understood because the mechanical properties of the skin at nanoscale are difficult to estimate, but the baseline sensor paradigm relies on MWCNTs penetrating the outer skin layer to a depth of 10-15 μm. Another possible enabling mechanism is capacitive coupling enhanced by the increase in surface contact area. Although MWCNTs can be coated for improved transduction, a polarizable uncoated prototype has first been tested and is reported here.

As the experimental prototype was designed to validate the performance of the dry MWCNT interface, the first MWCNT arrays were mounted on state of the art commercial active electrodes (i.e., with on-site amplification) and connected to commercial-off-the-shelf research electrophysiology recording equipment for comparison with existing "wet" sensors (Biosemi Active 2 **[9-10]**). The design and testing of a wireless sensor system with fully integrated MWCNTs and electronics will be reported elsewhere.

The electrode prototypes used for validation in all tests reported here had the following interface characteristics: MWCNT arrays with a nanotube diameter of *ca.* 50 nm, MWCNT length of 10-15 μm, Fe catalyst and no coating. The MWCNT arrays were grown on highly doped silicon substrates (Charntec Electronics <1-0-0> N type 0.8 - 0.15 $\Omega$.cm) using plasma enhanced chemical vapour deposition (PECVD) of acetylene over an iron catalyst **[6]**. Generally, a *ca.* 10 nm iron film was sputtered onto the silicon wafer immediately after etching the native oxide by immersion in hydrogen fluoride. The growth process consisted of heating the substrate to 650 ºC for 20 min in vacuum in order to break the Fe film into small islands between 50 - 100 nm in diameter. During the growth process the substrate was maintained at 650 ºC and acetylene was introduced to the chamber at a concentration of 5.0 % with $H_2$ as the carrier gas.

Prior to the human trials, a series of some simple tests were carried out as a quality check of the electrodes. These tests are described fully in **[11]** and we only provide an overview here. A first requirement for an electrode is that it exhibit low noise in the region of 1-2 μV RMS or less in the ~0.1-100 Hz range when immersed in saline solution. We carried out this test both for the prototype electrodes and the commercial electrodes in saline solution. The noise measured by the new electrodes is low and rather similar to that of the commercial electrodes **[11]**.

In order to test signal response in a semi-realistic situation, the electrodes were placed on pig skin and a small test signal applied beneath the skin. Pig skin is similar in structure to human skin and provides a good starting point for prototype development. The standard electrode was applied to pig skin using electrolytic gel while the new electrode was applied without gel or other skin preparation. The results from the comparison were again very similar.

For completeness, human tests were first carried out with conventional electrodes without gel and with prototype electrodes without CNT array interfaces (but with the same substrate). In both cases, the measurements were contaminated by a large amount of noise (1-2 orders of magnitude above those with wetted electrodes), as expected.

Due to ethical considerations, limited tests were conducted with one subject only (the tests were approved by the Ethical Committee of the University of Barcelona and the University of Surrey). The experimental protocol was designed to mimic typical EEG measurements conducted under both clinical and research practices. The sensor was thus evaluated under normal brain state conditions but also under experimental situations wherein characteristic brain responses are required. Hence, the protocol encompassed both spontaneous EEG and ERPs **[12]**.

EEG measurements were collected simultaneously again by a wet commercial electrode and the dry prototype electrode. The two electrodes were placed near each other in the scalp area (near the Fp2 position in the 10-20 system) and referenced to the tip of the nose. Eye movements (EOG) were recorded by two commercial wet electrodes placed at the outer canthus of each eye. A single wet ground electrode was placed on the cheek of the subject. EEG signals were sampled at 1024Hz with no band-pass on-line filtering. The participant sat in a comfortable chair in a dimly lit and electrically and acoustically shielded booth.

One of the well-known features in human EEG is the appearance of alpha waves (8-12 Hz) and the associated spectral peak when a subject relaxes, which is clearly visible when the subject closes the eyes. In the eyes-open condition, alpha EEG dominance is suppressed and an increase in the beta ($\beta$) range frequency (15-30Hz) can appear.

In order to test whether cortical alpha and beta oscillatory activity can be similarly recorded by the new electrode and the wet commercial electrodes, two protocols were employed. The first one was based on recording a long period of eyes-open and eyes-closed states, thus leading to the collection of long periods of alpha and beta oscillatory activity. The second one featured short time periods of alternating eyes open and closed condition states. The alternation of eyes opening and closing within brief time periods allows for the evaluation of the reactivity of brain rhythms to light stimulation.

Spontaneous EEG recordings consisted of two periods of 5 min during which the subject did not perform any task. In the eyes-open condition, the subject was instructed to focus the eyes on a fixed point on a computer screen, thus avoiding eye-blinks. In the eyes-closed condition, the subject was just required to keep the eyes closed and to relax.

EEG transition recordings consisted of a 3 minutes EEG recording with an alternation of 5 seconds of eyes-open and eyes-closed conditions. A pure tone (90 db SPL, 1000Hz and 10 ms rise/fall) was presented binaurally with headphones to advise the subject that period conditions had changed. The subject was required to open and close the eyes following the auditory tone.

ERPs are other well-established events in the study of cerebral electrical activity. ERPs are cerebral responses to the presentation of a sensory stimulus. The cerebral response is recorded as an EEG and is analyzed in the form of waves, characterized by their latency of appearance and amplitude.

Here, the appearance of an automatic auditory ERP component, the auditory N1, was evaluated. N1 is reflected by prominent scalp fronto-central negativity EEG deflection with a stimulus onset latency around 100-150ms. Auditory N1 reflects the transitory auditory cortical areas (temporal lobes) response to an auditory stimulus.

Aural stimulation was implemented by means of the stimulation software Presentation®

(Version 7.0) with the aid of portable equipment, presenting the visual stimuli (i.e., fixation point) and auditory stimuli through headphones. While the subject focused on a distracting task (reading), a sequence of auditory stimuli of two different kinds, namely a standard (p = 0.8; 75 ms; 90 dB SPL, 1000Hz and 10 ms rise/fall) and a deviant one (p = 0.2; 25 ms; 90 dB SPL, 1000Hz and 10 ms rise/fall) was presented binaurally with headphones. The sequence consisted of stimuli-trains formed first by a standard or deviated stimulus (at random) and followed by two standard stimuli. The interval between the stimuli within trains was 300 ms, while the interval between the trains was 400 ms.

The spontaneous EEG recording power spectral density (PSD) was evaluated using Welch's method for each condition separately. The PSD in open and closed eye conditions can be observed in Figure 3 for each electrode type, revealing a high similarity of the EEG data collected with each electrode. Moreover, the appearance of an increased alpha frequency peak (8-12Hz) appearance in the eyes-closed condition as compared with eyes-open one for each electrode type can be clearly observed. Furthermore, a small but clear beta frequency peak (around 20Hz) can be observed in each electrode in the eyes-open condition.

With equivalent results, the periods of 5 second recordings for eyes-open and closed conditions were separated from the continuous EEG recording and their PSDs were computed. Subsequently, the PSDs of each 5 second trail were averaged within conditions and separately for each electrode. Alpha activity was detected within periods of 5 second EEG recordings for both electrodes, thus reinforcing the consistency of the results from the data collected by both electrodes.
ERPs were obtained off-line by averaging 400 ms EEG epochs, which included 100 ms pre-stimulus as a baseline period. A 2-30Hz digital band-pass filter was used just for ERP extraction. Due to the fact that EEG electrodes were placed near the Fp2 position, eye movements interfered strongly with brain electrical activity recording. Hence, a strict artefact rejection process was conducted in the EEG trial selection. Epochs were excluded if signal values exceeded ±100 μV in EEG or ±10 μV in EOG. Lastly, a visual inspection of the remaining trials was conducted by two experienced researchers. After artefact removal, a total of 67 trials were selected for further analysis.

Figure 4A shows all selected trials recorded by conventional and prototype electrodes. A large negativity between 100-200 ms from stimulus onset—the N1 ERP response at each electrode—can be observed with each electrode. The repeated measure *t*-test comparison between electrodes of all single trials through each time-frame recorded produced no significant differences through the 400 ms time period (P>0.01). Further, a prominent negativity at 100-200 ms can be observed from the all-trial averages, which reflects the auditory N1 ERP (Figure 4B).

The design of a CNT-based electrophysiology electrode is a fascinating and challenging multi-disciplinary exercise involving analysis of requirements, skin and CNT mechanical and electrical properties at nanoscale, electrochemistry and biocompatibility issues. In a previous publication, we analyzed requirements for EEG/ECG/EOG applications and provided the logic for the electrode design, starting from a careful analysis of all the noise sources and measurement requirements and we proposed a of dry electrode based on the use of MWCNTs to penetrate the outer skin cell layers and thus reduce the noise of measurements. The test results reported here provide robust evidence that this dry electrode performs rather similarly to state of the art research-oriented wet electrodes, thanks to an interface exploiting the control of surface features at nanoscales and the excellent properties of CNTs.

It is also important to note that the subject did not report any pain or special sensations on application of the CNT-coated electrode, and six months after the trials no side effects have appeared (irritation, redness, itching or else). Nevertheless, as with other nanotechnology research, further work is needed to explore safety issues.

## ACKNOWLEDGEMENTS

The CNT electrode prototype (ENOBIO) project has been partly funded by Starlab, and developed under the FP6 European Project SENSATION (FP6-507231) and with support of the Catalan NECOM grant SGR2005-00831.

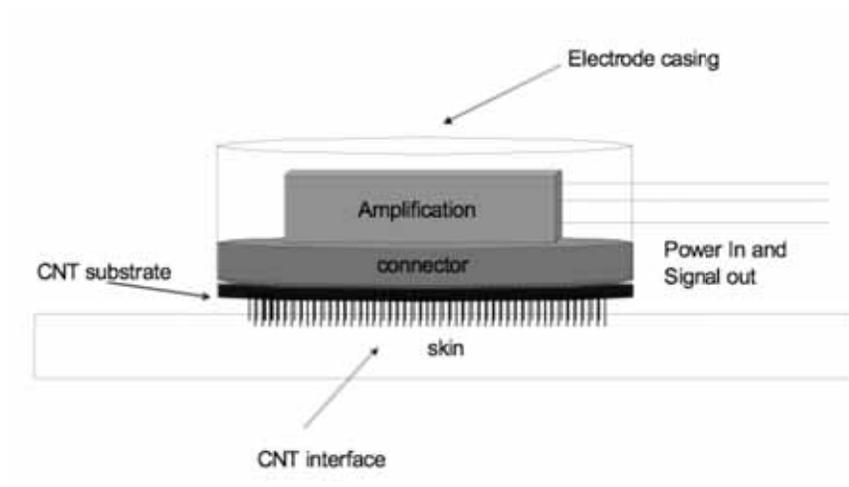

Fig. 1. Electrode prototype design: a CNT array grown on a conductive substrate is in contact with the Stratum Corneum. The signals are locally amplified for downstream sampling.

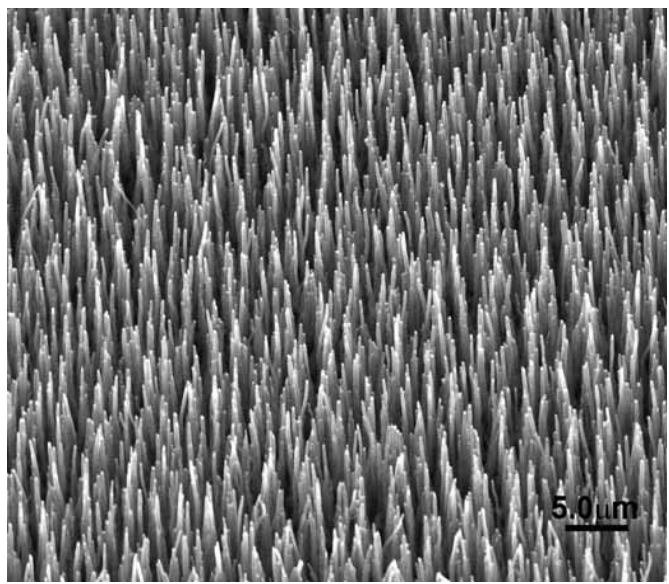

Fig. 2. Electron microscope image of the MWCNT array.

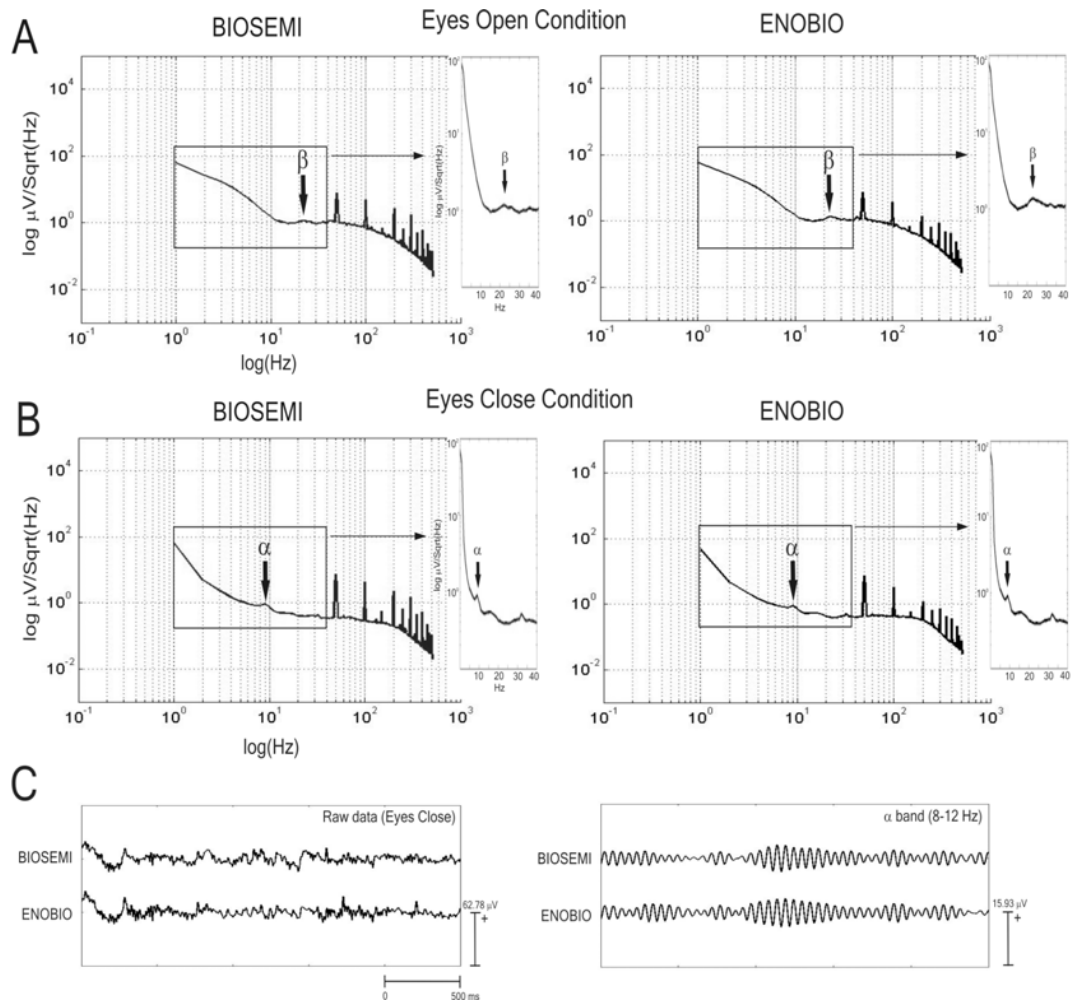

Figure 3. PSD results for each electrode (Biosemi and ENOBIO) for each EEG state condition. Beta (green arrow) and alpha (red arrow) frequency peaks are clearly observable in each condition. A large 50 Hz amplitude peak and its harmonics corresponding to the notch noise can be observed in both. A sample of eyes-closed raw data (with a bandpass of 1-35 Hz) and filtered to 8-12 Hz is presented at the bottom.

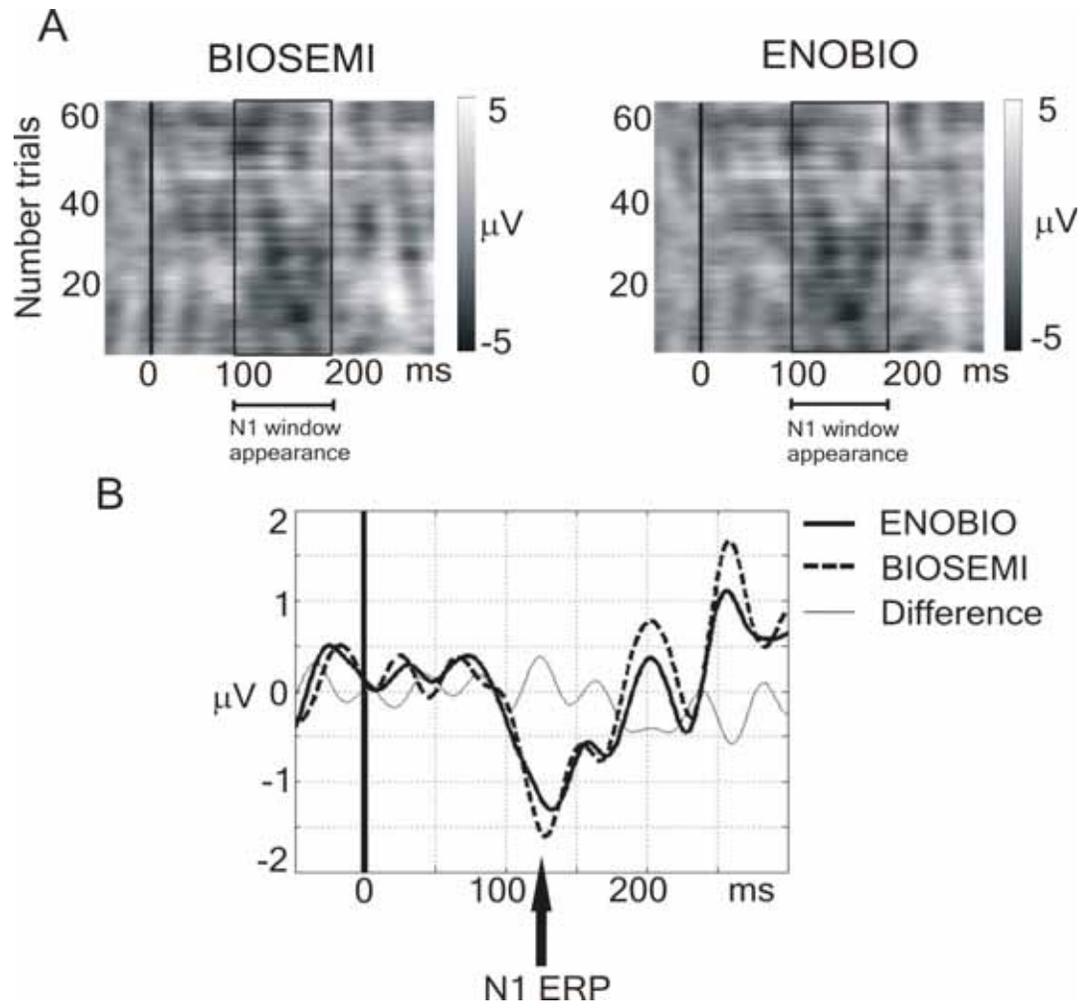

Figure 4. (A) Biosemi and ENOBIO representation of all 67 trials selected, and (B) results after averaging them. A clear auditory N1 event related potential is observed for both electrodes with a high similarity.